# DESIGN AND IMPLEMENTATION OF THE ADVANCED CLOUD PRIVACY THREAT MODELING

Ali Gholami[1], Anna-Sara Lind[2], Jane Reichel[2], Jan-Eric Litton[3], Ake Edlund[1], Erwin Laure[1]

[1]High Performance Computing and Visualization Department, KTH Royal Institute of Technology, Stockholm, Sweden
[2]Faculty of Law and Centre for Research Ethics and Bioethics, Uppsala University, Sweden
[3]Department of Medical Epidemiology and Biostatistics, Karolinska Institutet, Sweden

## ABSTRACT

*Privacy-preservation for sensitive data has become a challenging issue in cloud computing. Threat modeling as a part of requirements engineering in secure software development provides a structured approach for identifying attacks and proposing countermeasures against the exploitation of vulnerabilities in a system. This paper describes an extension of Cloud Privacy Threat Modeling (CPTM) methodology for privacy threat modeling in relation to processing sensitive data in cloud computing environments. It describes the modeling methodology that involved applying Method Engineering to specify characteristics of a cloud privacy threat modeling methodology, different steps in the proposed methodology and corresponding products. In addition, a case study has been implemented as a proof of concept to demonstrate the usability of the proposed methodology. We believe that the extended methodology facilitates the application of a privacy-preserving cloud software development approach from requirements engineering to design.*

## 1. INTRODUCTION

Many organizations that handle sensitive information are considering using cloud computing as it provides easily scalable resources and significant economic benefits in the form of reduced operational costs. However, it can be complicated to correctly identify the relevant privacy requirements for processing sensitive data in cloud computing environments due to the range of privacy legislation and regulations that exist. Some examples of such legislation are the EU Data Protection Directive (DPD) [1] and the US Health Insurance Portability and Accountability Act (HIPAA) [2], both of which demand privacy-preservation for handling personally identifiable information. Further than that, there is no consensus within International law whether specific requirements should be applicable to genetic information. There are several documents at the regional and international level that include some guidelines, for example UNESCO International Declaration on Human Genetic Data (2003) and the OECD Guidelines on Human Biobanks and Genetic Research Databases (2009). The Council of Europe has enacted an Additional Protocol to the Convention on Human Rights and Biomedicine, concerning Genetic Testing for Health Purposes (2008), which is legally binding to the states that has ratified it.

Also national law might contain specific regulation entailing further conditions and criteria for the handling to be legal and thereby legitimate. In Sweden for example, the Act on Genetic Integrity[1] lays down specific requirements when performing genetic testing or taking part of the genetic information. These type of requirements also exist in other Member States.

Threat modeling is an important part of the process of developing secure software – it provides a structured approach that can be used to identify attacks and to propose countermeasures to prevent vulnerabilities in a system from being exploited [3]. However, the issues of privacy and security are really two distinct topics [4] as security is a core privacy concept, and the current focus of the existing threat modeling methodologies is not on privacy in cloud computing, which makes it difficult to apply these methodologies to developing privacy-preserving software in the context of cloud computing environments.

In 2013, the Cloud Privacy Threat Modeling (CPTM) [6] methodology was proposed as a new threat modeling methodology for cloud computing. The CPTM approach was originally designed to support only the EU DPD, for reducing the complexity of privacy threat modeling. Additionally, there were weaknesses in threat identification step through architectural designs in the early stages of Software Development Life Cycle (SDLC) that demanded improvements.
This paper describes an extension of the CPTM methodology according to the principles of Method Engineering (ME) [5]. The method that has been applied is one known as "Extension-based", which is used for enhancing the process of identifying privacy threats by applying meta-models/patterns and predefined requirements. This new methodology that is being proposed provides strong methodological support for privacy legislation and regulation in cloud computing environments. We describe the high-level requirements for an ideal privacy threat modeling methodology in cloud computing, and construct an extension of CPTM by applying the requirements that were identified. In addition, we provide a case study [6] containing 8 key privacy requirements and 26 threats according to the DPD.

The rest of this paper is organized as follows. Section 2 provides a background to these developments by outlining the CPTM methodology and existing related work. Section 3 describes the characteristics that are desirable in privacy threat modeling for cloud computing environments. Section 4 describes the steps and products for the proposed new methodology. Section 5 describe a case study and implements the proposed design. Section 6 presents the conclusions from this research and directions for future research.

## 2. BACKGROUND AND RELATED WORK

The CPTM [6] methodology was proposed as a specific privacy-preservation threat modeling methodology for cloud computing environments that process sensitive data within the EU's jurisdiction. The key differences between the CPTM methodology and other existing threat modeling methodologies are that CPTM provides a lightweight methodology as it encompasses definitions of the relevant DPD [1] requirements, and in addition that it incorporates classification of important privacy threats, and provides countermeasures for any threats that are identified. In [22] the authors implemented the CPTM as a proof of concept for a data-intensive genomics cloud platform [21] and identified the weaknesses to be refined in this paper.

For the first step in the CPTM approach, the DPD terminology is used to identify the main entities to cloud environments that are in the process of being developed. Secondly, the CPTM methodology describes the privacy requirements that must be implemented in the environment, e.g., lawfulness, informed consent, purpose binding, data minimization, data accuracy, transparency, data security, and accountability. Finally, the CPTM approach provides countermeasures for the identified threats. Detailed description of the CPTM methodology steps have been discussed in [6] (Sections 3, 4 and 5).

---

[1] Lag (2006:351) om genetisk integritet m.m.

While the CPTM methodology was the first initiative for privacy threat modeling for cloud computing environments in accordance with the EU's DPD, it nevertheless does not support other privacy legislation, such as that required under the HIPAA [2]. In this paper, we identify the CPTM methodology weaknesses in supporting different privacy legislation and threat identification process and refine the methodology by applying an Extension-based ME approach. There has been a significant amount of research in the area of threat modeling for various information systems with the goal of identifying a set of generic security threats [7], [8], and [9]. There are guidelines for reducing the security risks associated with cloud services, but none of these include an outline of privacy threat modeling. The Cloud Security Alliance (CSA) guidelines [10] are not thorough enough to be referred as a privacy threat model because they are not specific to privacy-preservation. In [19, 20] authors discuss the security and privacy issues of processing sensitive big data in cloud computing environments, including loss of control, multi-tenancy, and lack of trust, in addition to survey privacy-preservation data processing approaches. The European Network and Information Security Agency (ENISA) has identified a broad range of both security risks and benefits associated with cloud computing, including the protection of sensitive data [11]. Pearson [4] describes the key privacy challenges in cloud computing that arise from a lack of user control, a lack of training and expertise, unauthorized secondary usage, complexity of regulatory compliance, trans-border data flow restrictions, and litigation.

LINDDUN [12] is an approach to privacy modeling that is short for "likability, identifiability, non-repudiation, detectability, information disclosure, content unawareness, and non-compliance". This approach proposes a comprehensive generic methodology for the elicitation of privacy requirement through mapping initial data flow diagrams of application scenarios to the corresponding threats. The Commission on Information Technology and Liberties (CNIL) has proposed a methodology for privacy risk management [13] that may be used by information systems that must comply with the DPD.

# 3. CHARACTERISTICS OF A PRIVACY THREAT MODELING METHODOLOGY FOR CLOUD COMPUTING

This section describes the features that we believe a privacy threat model should have in order to be used for developing privacy-preserving software in clouds in an efficient manner. Based on the properties that are identified, we then apply the Extension-based methodology design approach to construct an extension of the CPTM for supporting various privacy legislation in Section 4.

## 3.1. Privacy Legislation Support

Methodological support for the regulatory frameworks that define privacy requirements for processing personal or sensitive data is a key concern. Privacy legislation and regulations can become complicated for cloud customers and software engineering teams, particularly because of the different terminologies in use in the IT and legal fields. In addition, privacy threat modeling methodologies are not emphasized in existing threat modeling methodologies, which causes ambiguity for privacy threat identification.

## 3.2. Technical Deployment and Service Models

Cloud computing delivers computing software, platforms and infrastructures as services based on pay-as-you-go models. Cloud service models can be deployed for on-demand storage and computing power can be provided in the form of software-as-a-service (SaaS), platform-as-a-service (PaaS) or infrastructure-as-a-service (IaaS) [14]. Cloud services can be delivered to consumers using different cloud deployment models: private cloud, community cloud, public cloud, and hybrid cloud. The five essential characteristics of cloud computing are defined as on-demand self-service, broad network access, resource pooling, rapid elasticity, and measured service [14].

### 3.3. Customer Needs

The actual needs of the cloud consumers must be taken into consideration throughout the whole life cycle of a project. Additionally, during the course of a project, requests for changes often arise and these may affect the design of the final system. Consequently it is important to identify any privacy threats arising from the customer needs that result from such change requests. Customer satisfaction can be achieved through engaging customers from the early stages of threat modeling so that the resulting system satisfies the customer's needs while maintaining adequate levels of privacy.

### 3.4. Usability

Cloud-based tools aim at reducing IT costs and supporting faster release cycles of high quality software. Threat modeling mechanisms for cloud environments should therefore be compatible with the typical fast pace of software development in clouds-based projects. However producing easy-to-use products with an appropriate balance between maintaining the required levels of privacy while satisfying the consumer's demands can be challenging when it comes to cloud environments.

### 3.5. Traceability

Each potential threat that is identified should be documented accurately and be traceable in conjunction with the associated privacy requirements. If threats can be traced in this manner, it means that threat modeling activities are efficient in tracing of the original privacy requirements that are included in the contextual information and changes over the post-requirement steps such as design, implementation, verification and validation.

## 4. METHODOLOGY STEPS AND THEIR PRODUCTS

Motivated by the facts that privacy and security are two distinct topics and that no single methodology could fit all possible software development activities, we apply ME that aims to construct methodologies to satisfy the demands of specific organizations or projects [17]. In [5], ME is defined as "the engineering discipline to design, construct, and adapt methods, techniques and tools for the development of information systems".

There are several approaches to ME [17, 15] such as a fundamentally "ad-hoc" approach where a new method is constructed from scratch, "paradigm-based" approaches where an existing meta-model is instantiated, abstracted or adapted to achieve the target methodology, "Extension-based" approaches that aim to enhance an existing methodology with new concepts and features, and "assembly-based" approaches where a methodology is constructed by assembling method fragments within a repository.

Figure 1 represents different phases in a common SDLC. Initial security requirements are collected and managed in the requirements engineering phase (A). This includes identifying the quality attributes of the project and assessing the risk associated with achieving them. A design is composed of architectural solution, attack surface analysis and the privacy threat model. Potential privacy threats against the software that is being developed are identified and solutions are proposed to mitigate for adversarial attacks (B). The proposed solution from the design phase is implemented through a technical solution and deployment (C). This includes performing static analysis on source code for software comprehension without actually executing programs. The verification process (D) includes extensive testing, dynamic analysis on the executing programs on virtual resources and fuzzing as a black-box testing approach to discover coding errors and security loopholes in the cloud system. Finally, in the Validation phase the end-users participate to assess the actual results versus their expectations, and may put forth further change requests if needed.

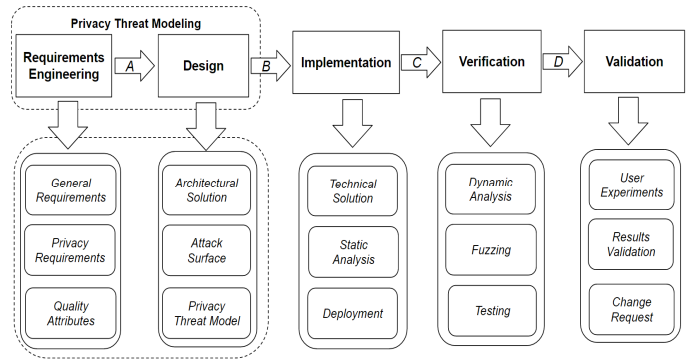

Figure 1, Privacy Threat Modeling in Requirements Engineering and Design of a SDLC

Our proposed methodology identifies the privacy requirements in the Requirements Engineering step, as shown in Figure 2. The results from the Requirements Engineering, which include specifications for privacy regulatory compliance, are fed into the Design step, where activities such as specifying the appropriate cloud environment, identifying privacy threats, evaluating risks and mitigating threats are conducted. Then the produced privacy threat model would be used in the implementation step finally it would be verified and validated in the subsequent steps.

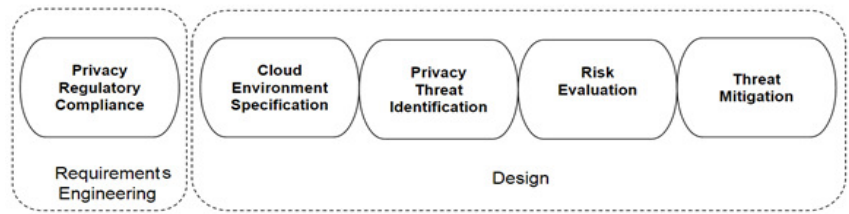

Figure 2, Overview of the Extended CPTM Methodology Steps

Cloud stakeholders and participants such as cloud users, software engineering team and legal experts will engage in the activities shown in Figure 2 to implement the threat model in context of steps A and B in Figure 1. Cloud software architect as a member of the software engineering team initiates a learning session to clarify the methodology steps and their products, privacy requirements (introducing the law title that is needed to be enforced in the cloud environment), and quality attributes such as performance, usability. The legal experts will identify the definitive requirements that ensure the privacy of data in the platform. In the Design step, the cloud software architect presents architecture of the developing cloud environment for various participants. This will result in a unified terminology to be used in the privacy threat model. The rest of this section outlines the implementation model of the steps represented in Figure 2.

## 4.1. Privacy Regulatory Compliance

Interpreting privacy regulatory frameworks can often be complex for software engineering teams. Further, the law is not constant since both legislators and courts may enact and interpret decisions that change and affect the law as it stands. For example, the verdict of the Court of Justice of the European Union in October 2015 to invalidate the Safe Harbor Agreement may have rather important repercussions for transferring data between the EU and the USA [24]. In the privacy regulatory compliance step, learning sessions with privacy experts, clients (cloud consumers) and requirements engineers facilitates the elicitation of privacy requirements (PR). For example, in the EU's DPD some of the privacy requirements are: lawfulness, informed consent, purpose binding, transparency, data minimization, data accuracy, data security, and accountability [6]. These principles are also included in the General Data Protection Regulation which the EU legislator is expected to enact during spring 2016 [26]. Each of the requirements that are identified will be labeled with an identifier, e.g., (PR$i$), name and description to be used in later stages.

## 4.2. Cloud Environment Specification

To ensure that the final cloud software will comply with the relevant legal and regulatory framework, several of the key characteristics that are affected by cloud computing services (including virtualization, outsourcing, offshoring, and autonomic technologies) must be specified. For this purpose, the physical/logical architectures of the deployment and service model can be developed according to the following steps.

- **Step A:** Define the cloud actors [18] (such as ***Cloud Consumer, Cloud Provider***, ***Cloud Auditor***, ***Cloud Broker***, and ***Cloud Carrier***). Cloud consumer is a person or organization uses service from cloud providers in context of a business relationship. Cloud provider makes service available to interested users. Cloud auditor conducts independent assessment of cloud services, operations, performance and security of the deployment. Cloud broker manages the use, performance and delivery of cloud services and establishes relationships between cloud providers and cloud consumers. Cloud carrier provides connectivity and transport of cloud services from cloud providers to cloud consumers through the network.
- **Step B:** Describe a detailed model of the cloud deployment physical architecture where the components will be deployed across the cloud infrastructure. This should give details of where the components will be deployed and run, for example, the operating system version, the database version, the virtual machine location, and where the database server will run.
- **Step C:** Describe the logical architecture of the cloud services model where the major cloud services, along with and the relationships between them that are necessary to fulfill the project requirements, are recorded. This should include the data flow and connections between the relevant cloud services and actors. Note that in this context, an entity is a cloud service with a set of properties that meet a specific functional requirement.
- **Step D:** Describe the assets that need to be protected, the boundaries of the cloud and any potential attackers that might endanger either the cloud environment or the assets that have been identified as being associated with that particular cloud.

The cloud environment specification step consists of composing an architectural report including assets that are subject to privacy protection, cloud actors, physical architecture of the deployment model, and logical architecture of the service model.

## 4.3. Privacy Threat Identification

In this step, privacy threats against the PRs that were established in section 3.1 will be identified and analyzed. To achieve this, the system designers will undertake the following steps.

- **Step A:** Select a privacy requirement from the PR list for threat analysis, e.g., (PR2).
- **Step B:** Correlate identified cloud actors (Step A from Section 3.2) with the actor roles that are defined in the project's privacy law. For example, correlating the **Data Controller** role as a **Cloud Consumer**, or the **Data Processor** role as a **Cloud Provider** in the DPD.

  **Step C:** Identify the technical threats that can be launched by an adversary to privacy and label them in the specified cloud environment. The cloud environment specification in Section 4.2 facilitates the threat identification. In SaaS clouds only application-level threats are required to be addressed. While in PaaS clouds threat identification focuses on OS and application level, where in IaaS clouds emphasis is physical hardware, network and storage. The attack surface for each service model varies with the cloud deployment model. For example, privacy threats within a private cloud might be smaller than a public cloud that is run by a commercial vendor. Public, hybrid or community clouds pose specific privacy threats on data locality which can be considered as an important privacy requirement.

  Each identified threat can be named as a T$i$,$j$, where $i$ indicates that threat T that corresponds to PR$i$ and $j$ indicates the actual threat number. For example, in T$_{2,5}$ $2$ indicates relevance of the threat to PR$2$ and $5$ is the actual threat number.
- **Step D:** Repeat the previous steps until all PRs are processed.

The threat identification step consists of composing an analysis report including a list of threats including id, name, date, author, threat scenario for each class of the PRs.

## 4.4. Risk Evaluation

In this step, project actors participate to rank the threats that have been identified in Section 3.3 with regard to their estimated level of importance and the expected severity of their effect on the overall privacy of the cloud environment. The **Importance (I)** indicates the likelihood of a particular threat occurring and the level of the **Effect (E)** indicates the likely severity of the damage if that threat against the cloud environment were carried out.

This step results in composing a risk evaluation report similar to the case study in Table 3 (Section 5.5). This report prioritizes the importance and effects of the privacy threats and it will be used in the Threat Mitigation step in Section 3.5.

## 4.5. Threat Mitigation

In this step, the threat modeling team proposes countermeasures to the threats that were identified in the previous step as having the highest likelihood of occurrence and the worst potential effects on the cloud environment. Each countermeasure should clearly describe a solution that reduces the probability of the threat occurring and that also reduces the negative effects on the cloud if the threat was carried out.

Finally, the recommended countermeasures from this step should be documented and fed into the implementation step to be realized through coding and for their effectiveness to be assessed by static analysis. In the later stages of verification and validation, each such countermeasure will be evaluated and approved by the participants.

# 5. CASE STUDY

Presented here is a description of the implementation of our proposed methodology for a PaaS BiobankCloud [21]. This includes a high-level definition of the BiobankCloud and an exemplary workflow. We follow the guidelines from Section 4 and demonstrate products of each step in the methodology.

## 5.1. Overview

The BiobankCloud platform is a collaborative project bringing together computer scientists, bioinformaticians, pathologists, and biobankers [21]. BiobankCloud aims to provide the capability of deploying sequencing applications with their dependencies within an environment called a container within a cloud computing environment.

Assume the BiobankCloud participants are Alice (Researcher) as Cloud Consumer representative, Bob (BiobankCloud) as Cloud Provider, Dennis as Auditor, Tom as Lawyer and Ove as Cloud Software Architect evaluate the corresponding risk of each identified threats. A typical scenario to use the BiobankCloud platform by Alice (Researcher) is following steps.

1. Alice registers an account in the platform and automatically will be considered as a guest user.
2. Bob enables Alice and entitles appropriate access control levels.
3. Alice opens the login page and enters the authentication credentials.
4. Bob authenticates the user according the presented credentials.
5. If the Alice is authenticated, Bob will permit her login to the platform.
6. Alice uploads genomics data and tries to run a workflow in the platform.
7. Bob ensures that Alice has enough permission to run the workflow.
8. Depending on the permission check:
   a) If Alice is not authorized to run the workflow she will be denied by Bob.
   b) If Alice is granted to access the resources demanded by the workflow, she will be permitted
9. Bob authorizes Alice to run the workflow and access the genomic data in the platform through the platform execution cluster. This cluster stores the results in the platform and presents the results to Alice.
10. The platform stores the log information from the above steps for auditing purposes by Dennis (Auditor).

This usage scenario will be used in the following steps (5.3 and 5.4) for actor-role correlations.

## 5.2. Privacy Requirements

The BiobankCloud platform demands to enforce the DPD requirements. The DPD is the EU's initial attempt at privacy protection, containing 72 recitals and 34 articles to harmonize the regulations for information flow within the EU Member States. The identified privacy requirements include the main roles of the DPD and the DPD fundamental PRs are lawfulness, informed consent, purpose binding, data minimization, data accuracy, transparency, data security, and accountability [6].

The DPD also highlights the demand for cross-border transfer of data through non-legislative measures and self-control. One example of where these types of privacy principles are being used is the Safe Harbor Agreement (SHA) which made it possible transfer data to US-based cloud providers that are assumed to have appropriate data protection mechanisms. However, as seen above, the European Court of Justice declared the SHA invalid in a ruling in October 2016 [25].

There is an ongoing effort [23, 28] to replace the EU DPD with a new a General Data Protection. Regulation containing more than 130 recitals and 91 articles that aim to lay out a data protection framework in Europe [26]. The proposed regulation will have a wider scope of application, covering processing of data from third state that is directed to the EU, such as to offer goods and services. Also the understanding of what is personal data is expanded in one specific context, namely through recital 25a of the proposal of the Regulation where it is stated that genetic data should be defined as personal data relating to the genetic characteristics of an individual which have been inherited or acquired as they result from an analysis of a biological sample from the individual in question, in particular DNA or RNA. This will restrict the processing of genetic data, since it could never be considered to fall outside the Regulation. The regulation also includes definitions of new roles related to handling data (such as data transfer officers) and considerably strengthens the role and functions of national data authorities [28]. For example, the regulation confers the power to the authorities to impose significant penalties for privacy breaches that result from violations of the regulations, for example, such a penalty could be 0.5 percent of the worldwide annual turnover of the offending enterprise [27].

The privacy requirements of the DPD including the DPD roles have been discussed in [6] and in the following we briefly summarize those identified requirements.

The main roles of the DPD participants are the data subject, controller, and processor. Article 2.a of the DPD defines data subject as an identifiable individual associated with the personal data. In Article 2.d, controller is defined as the natural or legal person, public authority, agency or any other body which alone or jointly with others determines the purpose and means of the processing of personal data. The processor acts as the natural or legal person, public authority, agency or any other body which processes personal data on behalf of the controller, as defined in Article 2.e.

**PR1 Lawfulness** sets out the basic premises for the legitimate processing of data, that all processing must be conducted within the regulatory framework of the DPD.

**PR2 Informed Consent** justifies processing of genomic data in the BiobankCloud. The genomic data may have been provided with informed consent through the data provider which constitutes the main justification for processing.

**PR3 Purpose Binding** ensures that personal data processing is performed according to predetermined purposes. The collected genetic data in the BiobankCloud will only be processed according to the purposes covered by the informed consent given by the subject or, if the law applicable to the Data Provider so admits, according to further purposes within the legal framework.

**PR4 Data Minimization** restricts extra and unnecessary disclosure of information to third parties, such as cloud provider, to reduce the risk of information leakage that leads to privacy breaches.

**PR5 Data Accuracy** describes the necessity to keep data accurate and to be updated by the Data Provider. A controller holding personal information shall not use that information without taking steps to ensure with reasonable certainty that the data are accurate and up to date.

**PR6 Transparency** entitles the data subjects to have information about the processing of their data and thereby a means to learn of the processing operation of their data. Transparency thus functions as a prerequisite for the data subjects to monitor that the data is accurate, in accordance to PR5.

**PR7 Data Security** proposes implementing technical measures to provide legitimate access and organizational safeguards. The Data Provider shall ensure that whoever processes the data on his behalf, e.g., the cloud provider assures adequate levels of security against unlawful data processing.

**PR8 Accountability** mandates internal, external auditing and control for various assurance reasons. The Data Provider is responsible to ensure compliant of supplied genomic data usage to the cloud providers.

## 5.3. Cloud Environment Specification

**Step A:** The BiobankCloud actors are identified through the general use cases. For example, from the usage scenario, Researcher, Guest, Data Provider, Administrator, and Auditor actors are identified as described in Table 1.

Table 1, Correlating the Domain Actors to the Cloud Actors

| Domain Actor | Cloud Actor | Description |
|---|---|---|
| Researcher, Guest, Data Provider | Cloud Consumer: A person or organization that uses service from the BiobankCloud | ● Researchers who are affiliated with the institutions that hold the genomic data. The researcher acts under the responsibility of the Data Provider, as described in this section. The guest researcher conducts experiments on subjects genomic data. ● The Guest is able to log in to the BiobankCloud and browse public content of the platform. ● Data Provider is the person responsible for the data and permitting The Guest can only access the resources and data in the BiobankCloud after permission has been granted by the Data Provider. |
| Administrator | Cloud Provider: The BiobankCloud platform that makes service available to interested users. | The administrator enables the users of the platform and installs new libraries. This actor can also assign/revoke roles from the users. |
| Auditor | Cloud Auditor: Conducts independent assessment of cloud services, operations, performance and security of the deployment. | The Auditor is able to see major identity and access management events, in addition to user authentication, authorization events to ensure compliance with the privacy legislation. |

**Step B: BiobankCloud Physical Architecture**

The BiobankCloud is designed as a PaaS to be easily installed on a private cloud using Karamel and Chef [2] [21]. The Chef recipes parametrize Vagrant [3] to create the virtualized clusters and services over 14.04 images using Hadoop[4] Open Platform-as-a-Service (Hops), as shown in GlassFish server version 4.0.1 over JDK 7.0 is used to run the big data lab information management system (LIMS). This server is confined between external and internal firewalls to separate it from the trusted private network behind the internal firewall. The platform users are able to access the web pages from the browsers in their computers through an external firewall, while platform administrator can only access the platform within the trusted network.

---

[2] http://www.karamel.io/

[3] https://docs.vagrantup.com/v2/virtualbox

[4] https://hadoop.apache.org/

Figure 3. Hops utilize the Hadoop version 2.x within a Hadoop Cluster (Hops-YARN) and Hops infrastructure containing MySQL 7.x cluster and Hops file system (Hops-FS).

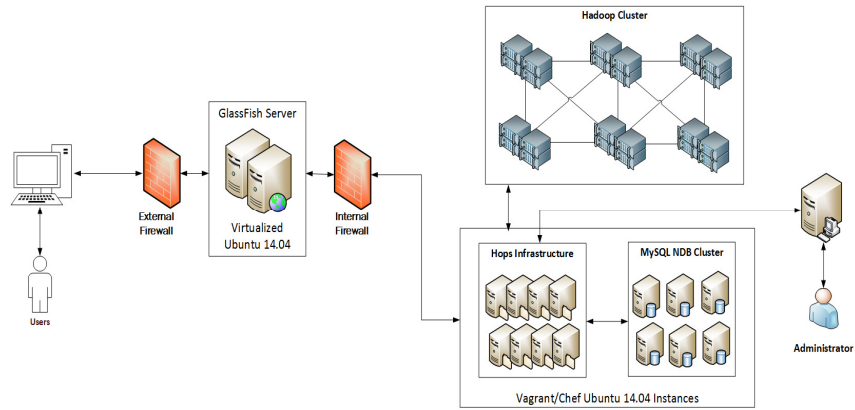

Figure 3: BiobankCloud Physical Architecture

**Step C: BiobankCloud Logical Architecture**

The BiobankCloud LIMS consists of several components for secure access, workflow execution, data sharing, as shown in Figure 4. Hops-FS as the next-generation of the Hadoop Filesystem (HDFS) supports multiple stateless Name Nodes but it keeps the metadata in a MySQL Cluster. Hops-YARN is another component that provides distributed stateless resource management with storing states in the MySQL Cluster for fault-tolerant.

The BiobankCloud services will be deployed in a PaaS private cloud. The user requests will be dispatched to a multi-tenant Hadoop cluster that is managed by the Hops-YARN. The Hadoop cluster runs submitted jobs and will notify the user through the LIMS interfaces. The Node Manager is an agent that is responsible for containers, monitoring their resource usage such as CPU, memory, network and reporting to the Hops-YARN resource manager.

A DataNode manages storage of each container node using keeping a set of blocks related to files. When a workflow is executed, its algorithmic dependencies will be fetched into the container as data-staging phase.

**Step D: Assets, Boundaries and Attackers**

The BiobankCloud platform stores genomic data which constitute as the main asset. The platform services that are provided to the external users can also be considered as assets. The security boundaries consist of firewalls that control the incoming/outgoing traffic through the BiobankCloud along with the physical means to deny access to the computing platform for unauthorized persons. Eavesdroppers and malicious users are known as attackers or adversaries that are able to exploit the possible vulnerabilities in the platform.

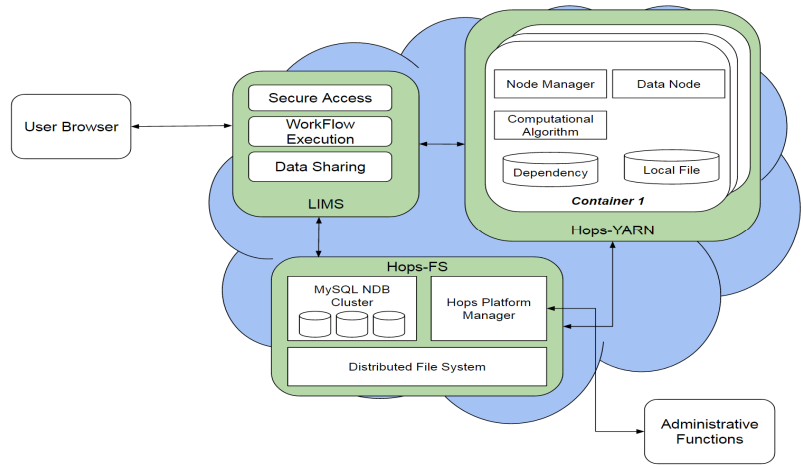

Figure 4: BiobankCloud Logical Architecture

## 5.4. Privacy Threat Identification

**Step A:** Select a privacy requirement from the PR list for threat analysis from {*PR1, PR2,…, PR8*}.

**Step B:** Correlate the domain actors to the DPD roles.

Table 2, Correlating the BiobankCloud actors with the DPD roles

| Cloud Domain Actor | DPD Role | Role Description |
|---|---|---|
| Cloud Consumer | Data Subject or Data Controller | • Data Subject: The individual person from which the genomic data has been extracted<br>• Data Controller: The entity that owns or provides the genomic data to the platform and decides on the processing of the data. |
| Cloud Provider | Data Processor | The BiobankCloud platform and all the administrative resources that provide the computing services to the Cloud Consumers. The Data Processer may only act on behalf of the Controller. |
| Cloud Auditor | Data Processor or Data Controller | Auditor inspects compliance with legal and accountable processing and storage of the genomic data according to the DPD rules. If the auditor simply processes the data on behalf of the controller (e.g, internal control) then he/she is a processor. Otherwise, he/she can act as a controller on his/her own right. |

**Step C:** Identify the technical threats that can be launched by an adversary to privacy and label them in the specified cloud environment. Each identified threat can be named as a $T_{i,j}$, where $i$ indicates that threat T that corresponds to PR$i$ and $j$ indicates the actual threat number. For example, in $T_{2,5}$ value *2* indicates relevance of the threat to PR*2* and value *5* is the actual threat number.

**T1 PR1 (Lawfulness)**

- $T_{1,1}$: Lack of relevant information on legal rights and duties to allow data subject and other interested parties to use effective means for accessing the data.
- $T_{1,2}$: Amendments of legal requirements or unawareness of new rules. Incorrect interpretation or application of legal concepts leading to unlawful processing of (complex) data.
- $T_{1,3}$: Lack of agreement from all entities regarding the processing the genomic data.
- $T_{1,4}$: If data are not obtained in accordance with the law, or without approval (informed consent or ethics board).

**T2 PR2 (Informed Consent)**

- $T_{2,1}$: Excessive terms of service (ToS) containing too much specific information that are not clear enough, e.g., legal terms that are not easily understandable for a layman.
- $T_{2,2}$: Lack of possibility to give consent dynamically to a specific subset of genomic data.

**T3 PR3 (Purpose Binding)**

- $T_{3,1}$: Researcher does not use the data according to the initial purpose or the Data Provider does not use the data according to the purposes.
- $T_{3,2}$: Researcher who has access to multiple data studies, makes cross-link analysis to the genomic data that is not consented and, hence illegal, according to the PR1 and PR2.

**T4 PR4 (Data Minimization)**

- $T_{4,1}$: The requested sensitive data to be used by cloud provider or guest researcher is not certain or well defined in advance.
- $T_{4,2}$: If the Data Provider does not define the retention period of the sensitive (genomic) data, there is a threat of accumulating more and more sensitive data over time, that can be used for inference and linking attacks.

**T5 PR5 (Data Accuracy)**

- $T_{5,1}$: If the Data Provider cannot or will not update data when having wrong information has been found to be incorrect.
- $T_{5,2}$: Data Provider uploads data to the cloud provider but data source validity is not affirmed.

**T6 PR6 (Transparency)**

- $T_{6,1}$: Lack of communication and information between entities. The threat is even more sever when the lack of information or openness is harmful for weaker parties, the sample donor/data subject.
- $T_{6,2}$: Researchers or Data Provider cannot get access to modify or erase data, due to unclear data processing procedures.

**T7 PR7 (Data Security)**

- $T_{7,1}$: Theft of authentication credentials by an adversary through phishing attacks or network eavesdropping, brute force attacks to guess authentication credentials or identities of users.

- $T_{7,2}$: Injecting code such as SQL injection, cross-site scripting (XSS) format, type or length of input data.
- $T_{7,3}$: Wide access to data by a large group of people.
- $T_{7,4}$: Elevation of privileges by an attacker to change the access rights to higher privileges.
- $T_{7,5}$: Unavailability of data due to denial of service (DoS) attacks.
- $T_{7,6}$: Session reply through message theft by eavesdropping to steal a session.
- $T_{7,7}$: Inference attacks through guessing or background knowledge to the minimized data, required by PR4.
- $T_{7,8}$: Storing passwords, credentials, database connections, keys in plain text or within the source code.
- $T_{7,9}$: Existing bugs in the OS kernel that can be target for an attacker to exploit them.

**T8 PR8 (Accountability)**

- $T_{8,1}$: Lack of mechanism for secure auditing to provide evidence of confidentiality and integrity.
- $T_{8,2}$: Logging information or audit trails contain sensitive information about the subjects.
- $T_{8,3}$: Excessive information in the audit logs to make audit and inspection about usage of the genomic data by an auditor.

## 5.5. Risk Evaluation

The BiobankCloud participants such as analyst, architects and lawyers participate and evaluate the important of the threats as summarized in the Table 3.

Table 3, Risk evaluation matrix for the identified threats. *I* indicates the likelihood of threat and *E* indicates the effect of exploiting the threat on the whole BiobankCloud.

| ID | Name | Exploit Scenario | *I* | *E* | Participants |
|---|---|---|---|---|---|
| $T_{1,1}$ | Lack of Lawfulness | An adversary uses the BiobankCloud to process or store genomics information due to lack of lawfulness information. | L | H | Bob, Dennis, Tom |
| $T_{1,2}$ | Amendments of Regulations | An adversary uses the BiobankCloud to avoid compliance with new regulations that may have strict privacy requirements. | L | M | Alice, Bob, Tom, Ove |
| $T_{1,3}$ | Lack of Agreement | An adversary is able to process and achieve results in the platform without agreeing to the BiobankCloud ToS. | M | H | Alice, Bob, Tom, Ove, Dennis |
| $T_{1,4}$ | Illegally Collected Data | An adversary unlawfully process data without ethical and legal permissions. | M | M | Alice, Bob, Tom, Ove, Dennis |
| $T_{2,1}$ | Excessive ToS | The BiobankCloud ToS contains many complicated legal terms that clients are not able to understand. | H | H | Alice, Bob, Tom, Ove |
| $T_{2,2}$ | Dynamic Consents | The BiobankCloud platform restricts user to update the consent and ethical information. | H | M | Alice, Bob, Tom, Ove |
| $T_{3,1}$ | Misuse of Data | Researcher does not use the data according to the initial purpose or the DP does not use the data | L | H | Alice, Bob, Ove, Dennis |
| | | according to the purposes. Also the Biobankcloud should not share the genomics data or users data to other external entities. | | | |
| $T_{3,2}$ | Cross-linking over Datasets | Researcher who has access to multiple data studies, makes cross-link analysis to the genomic data that is not consented. | M | H | Alice, Bob, Ove, Dennis |
| $T_{4,1}$ | Excessive Information | A Researcher uploads unnecessary information of data subjects that is not required by the platform. | M | L | Alice, Bob, Ove |
| $T_{4,2}$ | Data Minimization | The BiobankCloud platform accumulates huge amount of information over time from users who have not set data retention period. | M | M | Alice, Bob, Tom, Ove, Dennis |
| $T_{5,1}$ | Data Modification | The BiobankCloud is not able to update the user information or genomics data when Data Provider or Cloud Consumer decides to update the information. | M | M | Alice, Bob, Tom, Ove, Dennis |
| $T_{5,2}$ | Data Source Validity | An adversary uploads compromised genomics data without valid data source references. | L | L | Alice, Bob, Tom, Dennis |
| $T_{6,1}$ | Unclear Processing of Data | The BiobankCloud restricts other participants to access the information about the data usage in the platform and how genomic data is processed. | H | M | Alice, Bob, Tom, Ove |
| $T_{6,2}$ | Transparent Data Access | Researchers or Data Provider cannot get access to modify or erase data, due to unclear data processing procedures. | L | M | Alice, Bob, Tom, Ove, Dennis |
| $T_{7,1}$ | Weak Authentication | An adversary is able to steal user credentials through network eavesdropping or password theft and make reply attacks. | H | H | Alice, Bob, Tom, Ove, Dennis |
| $T_{7,2}$ | Input Validation | An adversary is able to issue SQL injection or XSS over web pages. | H | H | Alice, Bob, Ove |
| $T_{7,3}$ | Misusing the Roles | Wide access to data by a large group of people. | M | H | Alice, Bob, Tom, Ove, Dennis |
| $T_{7,4}$ | Unauthorized Access | An adversary is able to change his/her privileges in order to get permissions from other roles that he is not entitled. | M | H | Alice, Bob, Ove |
| $T_{7,5}$ | Unavailable Services | The BiobankCloud platform is not accessible due to technical or adversarial attacks. | M | M | Alice, Bob, Ove |
| $T_{7,6}$ | Session theft | An adversary is able to steal other users sessions to compromise integrity of data. | M | M | Alice, Bob, Ove |
| $T_{7,7}$ | Re-identification of Individuals | An adversary is able to re-identify individuals from the data sets with prior background knowledge. | M | H | Alice, Bob, Ove |

| $T_{7,8}$ | Theft of Credentials in Source Code | An adversary is able to steal the platform master keys and credentials that are written in the source code or unsecure text storage. | *M* | *H* | Alice, Bob, Ove |
|---|---|---|---|---|---|
| $T_{7,9}$ | Kernel Exploits | An attacker is able to exploit OS kernel bugs to gain access over the host VM or other guest VMs. | *H* | *H* | Alice, Bob, Ove |
| $T_{8,1}$ | Unsecure auditing | Auditor is not able to infer proof of CIA from the audit trails | *M* | *M* | Alice, Bob, Tom, Ove, Dennis |
| $T_{8,2}$ | Sensitive Info in Logs | An adversary is able to extract sensitive information such as data subjects personally identifiable information or credentials from the platform. | *L* | *M* | Bob, Tom, Ove, Dennis |
| $T_{8,3}$ | Unusable Log Files | Auditor is not able to extract information from usable audit reports. For example too much information or inconsistent log files due to time synchronization in the distributed systems. | *H* | *M* | Bob, Tom, Ove, Dennis |

## 5.6. Threat Mitigation

The cloud participants propose trade-off mitigation solutions as countermeasure for the evaluated threats from the previous step (Section 5.5) to meet the architectural goals such as privacy vs. usability. Each proposed countermeasure refers to the corresponding threat in Section 5.5, i.e. $C_{4,1}$ indicates countermeasure for the threat $T_{4,1}$.

$C_{1,1}$: Establish procedures with legal basis according to {*PR1, PR2,…, PR8*} to clarify the legitimate rights and responsibilities for each role.

$C_{1,2}$: Apply the possible changes from the new laws or regulations that might affect processing of the genomic data within the DPD framework.

$C_{1,3}$: Upon registration of new users provide a definitions of the platform ToS and user responsibilities.

$C_{1,4}$: Assume uploaded data could have been collected illegally by a Cloud Consumer and for this purpose ethical approval and consent forms of data subject must be presented and validated by Administrator role.

$C_{2,1}$: The BiobankCloud ToS should not contain complicated legal terms or conditions that are difficult for a non-law expert to interpret. It only requires to clearly describe the responsibilities short and precisely.

$C_{2,2}$: The consents and ethical information can be updated by the Cloud Consumer whenever there are new forms. The platform stores older versions for auditing purposes.

$C_{3,1}$: For every new data set there should be new consents and ethical approval forms. For example, if there is a running approved project and a set of new data is uploaded, the Administrator should be provided with further evidence of ethical permits over the study.

$C_{3,2}$: Restrict users with the aim to run cross-linking analytical jobs over multiple independent data sets that is not consented by the data subjects. Also data sets or Cloud Consumer information must be kept confidential to the BiobankCloud.

$C_{4,1}$: The platform should not be implemented in a way to require excessive information from the Researchers or data subjects.

$C_{4,2}$: Users will update the data retention period and when this date expires the Administrator manually delete the data after notifying the user.

$C_{5,1}$: Allow all users to be able to update their information on-demand in addition to update or erasure of genomics data.

$C_{5,2}$: Administrator should be able to approve or reject projects based on their validity. For example, if an uploaded data is incompatible with the platform or data has been acquired from unknown sources, the Administrator can reject processing that project.

$C_{6,1}$: The platform data usage should be transparent for authorized parties such as Administrators and Auditors to see all the activities. Data Provider to see the data access attempts on the uploaded data sets.

$C_{6,2}$: Provide functionalities to update the information such as genomics data or ethical approval consent forms that are required to be modified by the authorized roles. Updated information should be transparent to the user who made the changes.

$C_{7,1}$: Implement usable strong authentication mechanisms such as two-factor authentication, biometrics authentication or public key certificates to avoid compromising security of the platform through stealing user credentials.

$C_{7,2}$: Validate the input at least in the LIMS server to restrict any possible SQL injection, XSS or cookie/query string/HTTP manipulation.

$C_{7,3}$: The platform should classifies the responsibilities and permissions of each role to avoid granting a broad range of responsibilities to each user. For example restrict Auditor to have Administrator permissions.

$C_{7,4}$: For every attempt to access the resources ensure the identity of user and enforce the predefined polices for that user. For example in the LIMS platform ensure RBAC while in the Hops enforce discretionary access control in a fine-grained approach.

$C_{7,5}$: Protect the internal services through the internal firewall that restricts the I/O HTTPS traffic to another external firewall where is only accessible through a secure port. Using intrusion detection techniques identify the intruders and block their IP address.

$C_{7,6}$: Do not trust the safety of user computers and keep the session state information in the server side and do not send more than an opaque session identifier to the users. Encrypt whatever that is sent to the client with the server keys for making it confidential in the client machine over secure HTTPS channels. Also for every new session create new session information both in the server and for client.

$C_{7,7}$: Anonymize the micro data that contains explicit identifier of data subjects. For example, apply k-anonymity or l-diversity to generalize the sensitive attributes.

$C_{7,8}$: Avoid using the secret credentials such as connection names, passwords in the source code or storing them in plain text in the platform.

$C_{7,9}$: Run the Hops and other services in virtualized secure environments where integrity of images are ensured prior to installation. Apply latest patches to ensure safety of the VMs and ensure all communications between user and LIMS server are encrypted.

$C_{8,1}$: The platform should store log files related to study data, consents management, authentication, authorization, identity and access management for all users actions. Store the log files in the high availability Hops services. Do not modify or delete the log files.

$C_{8,2}$: Prevent the application, server, data bases components or the logger of the LIMS server to include sensitive credentials and information. For example default setting of Glassfish can be turned off to avoid printing the contents of issued queries.

$C_{8,3}$: Implement  custom audit modules for consistent logging of important feature such as browser, OS, IP/MAC addresses, timestamp, initiator of events, targets and outcome of the actions. This will ensure excessive information from the log files will not be displayed in the audit report. Additionally, the log server should be synchronized with the rest of process in the BiobankCloud to ensure time consistency in the audit reports.

# 6. CONCLUSIONS AND FUTURE WORK

In this paper we identified the requisite steps to build a privacy threat modeling methodology for cloud computing environments using an Extension-based Method Engineering approach. For this purpose, we extended the Cloud Privacy Threat Modeling (CPTM) methodology to incorporate compliance with various legal and regulatory frameworks, in addition to improving the threat identification process.

As a proof of concept, the proposed methodology was applied to a BiobankCloud that aims to process and store genomic data in a PaaS cloud environment. This BiobankCloud demands compliance with the DPD privacy requirements. We formulated the DPD privacy requirements and proposed a threat model by engaging the participants to provide a trade-off solution to meet the project architectural goals such as privacy vs. usability.

In future research, we aim to build a prototype using this methodology in context of the SDLC "implementation" step. Also, we would like to apply our methodology to other privacy regulations such as HIPAA to discover strengths and limitations of this approach.

## ACKNOWLEDGEMENTS

This work funded by the EU FP7 project Scalable, Secure Storage and Analysis of Biobank Data under Grant Agreement no. 317871.

## Authors

**Ali Gholami** is a PhD student at the KTH Royal Institute of Technology. His research interests include the use of data structures and algorithms to build adaptive data management systems. Another area of his research focuses on the security concerns associated with cloud computing. He is currently exploring strong and usable security factors to enable researchers to process sensitive data in the cloud.

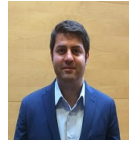

**Anna-Sara Lind** is Associate Professor and Senior Lecturer in Public Law at the Faculty of Law, Uppsala University. Her research focuses on public law, EU law and fundamental rights and on how EU law and international law interacts in the fields of Medical law and welfare state law in a complex constitutional reality. She is associated to the Centre for research ethics and bioethics and to the Faculty of Theology, Uppsala University. Lind is Deputy leader for the Impact of Religion-Challenges for Society, Law and Democracy.

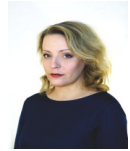

**Jane Reichel** is Professor of Administrative Law at the Faculty of Law, Uppsala University, and associated with the Centre for Research Ethics & Bioethics, Uppsala University. She is currently the chairman of the Research Committee at the Faculty of Law and the Vice Dean. Jane's research focuses on processes of globalization and Europeanisation of administrative law, especially within the area of administrative cooperation within research and biobanking, transparency and data protection.

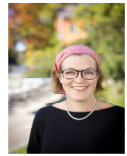

**Professor Erwin Laure** is Director of the PDC - Center for High Performance Computing Center at KTH, Stockholm. He is the Coordinator of the EC-funded "EPiGRAM" and "ExaFLOW" projects as well as of the HPC Centre of Excellence for Bio-molecular Research "BioExcel" and actively involved in major e-infrastructure projects (EGI, PRACE, EUDAT) as well as exascale computing projects. His research interests include programming environments, languages, compilers and runtime systems for parallel and distributed computing, with a focus on exascale computing.

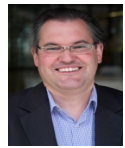